\documentclass{article}

\PassOptionsToPackage{numbers, compress}{natbib}

\usepackage[preprint]{neurips_2024}

\usepackage[utf8]{inputenc} 
\usepackage[T1]{fontenc}    
\usepackage{hyperref}       
\usepackage{url}            
\usepackage{booktabs}       
\usepackage{amsfonts}       
\usepackage{nicefrac}       
\usepackage{microtype}      
\usepackage[table]{xcolor}
\usepackage{wrapfig}
\usepackage{multirow}
\usepackage{multicol}
\usepackage{amsmath}
\usepackage{graphicx}
\usepackage{subfigure}
\usepackage[ruled,linesnumbered]{algorithm2e}

\usepackage{comment}
\usepackage{float}
\usepackage[export]{adjustbox}
\usepackage{caption}

\title{Explainable Multimodal Emotion Recognition}

\author{
	Zheng Lian$^{1}$, Haiyang Sun$^{1}$, Licai Sun$^{1}$, Hao Gu$^{1}$, Zhuofan Wen$^{1}$, Siyuan Zhang$^{1}$, \\
	\textbf{Shun Chen$^{1}$, Mingyu Xu$^{1}$, Ke Xu$^{1}$, Kang Chen$^{1}$, Lan Chen$^{1}$, Shan Liang$^{3}$,} \\
	\textbf{Ya Li$^{4}$, Jiangyan Yi$^{1,2}$, Bin Liu$^{1,2}$, Jianhua Tao$^{5,6}$} \\
	$^1$Institute of Automation, Chinese Academy of Sciences \\
	$^2$School of Artificial Intelligence, University of Chinese Academy of Sciences\\
	$^3$Department of Intelligent Science, Xi'an Jiaotong-Liverpool University \\
	$^4$School of Artificial Intelligence, Beijing University of Posts and Telecommunications\\
	$^5$Department of Automation, Tsinghua University \\
	$^6$Beijing National Research Center for Information Science and Technology, Tsinghua University \\
	\texttt{lianzheng2016@ia.ac.cn} \\
}

\begin{document}

\maketitle

\begin{abstract}
	Multimodal emotion recognition is an important research topic in artificial intelligence, whose main goal is to integrate multimodal clues to identify human emotional states. Current works generally assume accurate labels for benchmark datasets and focus on developing more effective architectures. However, emotion annotation relies on subjective judgment. To obtain more reliable labels, existing datasets usually restrict the label space to some basic categories, then hire plenty of annotators and use majority voting to select the most likely label. However, this process may result in some correct but non-candidate or non-majority labels being ignored. To ensure reliability without ignoring subtle emotions, we propose a new task called ``\textbf{Explainable Multimodal Emotion Recognition (EMER)}''. Unlike traditional emotion recognition, EMER takes a step further by providing explanations for these predictions. Through this task, we can extract relatively reliable labels since each label has a certain basis. Meanwhile, we borrow large language models (LLMs) to disambiguate unimodal clues and generate more complete multimodal explanations. From them, \textbf{we can extract richer emotions in an open-vocabulary manner}. This paper presents our initial attempt at this task, including introducing a new dataset, establishing baselines, and defining evaluation metrics. In addition, EMER can serve as a benchmark task to evaluate the audio-video-text understanding performance of multimodal LLMs.
\end{abstract}

\section{Introduction}
Multimodal emotion recognition has experienced rapid development in recent years \cite{baltruvsaitis2018multimodal, lian2024merbench}. Current works predominantly revolve around two aspects: the collection of larger and more realistic datasets \cite{poria2019meld, yu2020ch} and the development of more effective architectures \cite{tsai2019multimodal, lian2021ctnet}. Despite promising progress, emotion recognition suffers from label ambiguity \cite{picard2000affective}. It arises due to the inherent subjectivity in the emotion annotation process, i.e., different annotators may assign distinct labels to the same video. Label ambiguity results in potentially unreliable labels of existing datasets, bringing obstacles to the systems developed on these datasets to meet requirements in practical applications.

To enhance label reliability, current works mainly focus on restricting the label space to reduce the annotation diversity \cite{busso2008iemocap, lian2023mer}, while increasing the number of annotators and using majority voting to determine the most likely label \cite{li2017reliable, mollahosseini2017affectnet}. However, this approach may exclude correct but non-candidate or non-dominant labels, resulting in inaccurate annotations.

To obtain reliable labels but not ignore subtle ones, we introduce a new task called ``Explainable Multimodal Emotion Recognition (EMER)''. Unlike traditional emotion prediction, EMER goes a step further and provides explanations for these predictions. In this way, the identified labels are more reliable because there is a corresponding basis. Meanwhile, with the reasoning capability of large language models (LLMs), we can disambiguate unimodal clues and generate more comprehensive multimodal descriptions with rich emotion categories.

Another motivation behind EMER is that emotions are related to multi-faceted clues, such as prosody \cite{el2011survey}, facial expressions \cite{li2020deep} (or micro-expressions \cite{ben2021video}), gestures \cite{noroozi2018survey} (or micro-gestures \cite{chen2023smg}), etc. Current works generally identify emotions from one or several aspects. Unlike existing works, EMER provides a common format for emotion-related tasks, aiming to integrate all clues to generate more accurate labels. Meanwhile, emotions are complex. Current datasets limit the label space to a few categories, causing annotators being unable to describe emotional states accurately. Differently, EMER does not limit the label space and can generate richer labels in an open-vocabulary manner.

\begin{figure}[t]
	\centering
	\includegraphics[width=\linewidth]{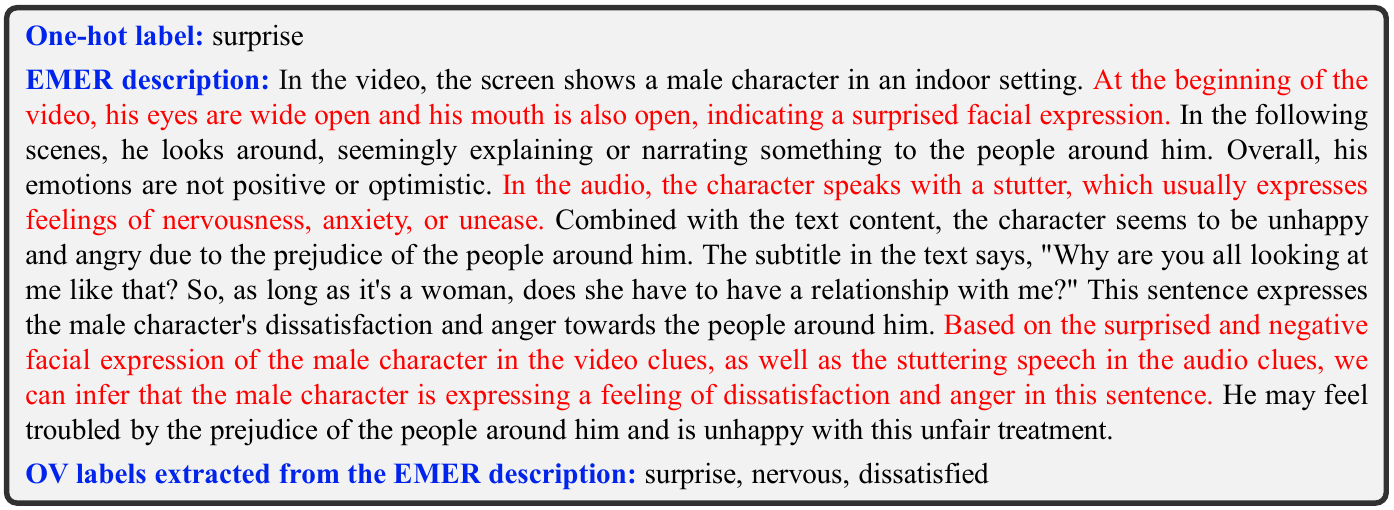}
	\caption{One example (``sample\_00000669'') to illustrate the differences between the one-hot label, EMER description, and EMER-based open vocabulary labels.}
	\label{Figure1}
\end{figure}

This paper proposes a new task EMER, aiming to achieve more reliable and accurate emotion recognition technology. To facilitate further research, we establish a new dataset, baselines, and evaluation metrics. Figure \ref{Figure1} shows the differences between the traditional one-hot label, EMER description, and EMER-based open vocabulary (OV) labels. We observe that more accurate labels can be extracted in this way. In addition to \emph{surprise}, we can also extract \emph{nervous} and \emph{dissatisfied}. The main contributions of this paper can be summarized as follows:
\begin{itemize}
	\item This paper introduces the EMER task for reliable and accurate emotion recognition. On the one hand, it provides the evidence and reasoning process for identified emotions. On the other hand, it can integrate all emotion-related clues to generate more accurate labels.
	
	\item To facilitate further research, we construct a dataset, establish baselines, and define evaluation metrics. Meanwhile, we will open-source the code and intermediate results.
	
	\item Besides emotion recognition, EMER can serve as a benchmark task to evaluate the audio-text-video understanding ability of multimodal LLMs (MLLMs).
\end{itemize}

\section{Related Work}
\label{sec:2}

\paragraph{Multimodal Emotion Recognition.} 
Multimodal emotion recognition aims to integrate multimodal clues to identify emotions. Unlike other tasks with clearly defined categories (such as object or action recognition), emotions are relatively ambiguous. Especially in multimodal scenarios, emotions are more complex \cite{busso2008iemocap} and there may be a modality repulsion problem \cite{yu2020ch} (i.e., different modalities may convey distinct emotions). To improve the annotation consistency, previous works often restricted the label space and used majority voting to determine the most likely label \cite{lian2023mer, li2017reliable}. For example, Lian et al. \cite{lian2023mer} employed at least six annotators and used multi-stage checks to select samples with explicit emotions. Li et al. \cite{li2017reliable} labeled each sample about 40 times and used the EM algorithm to filter out unreliable labels. Although these works enhance the label reliability, some correct but non-majority or non-candidate labels may be ignored. This paper introduces a new task, EMER, which provides a pathway to recognizing emotions in an open-vocabulary manner. With this task, we aim to generate more accurate labels for each sample. To the best of our knowledge, this is the first attempt to address emotion recognition in this manner.

\paragraph{Open Vocabulary Learning.}
Open vocabulary learning aims to identify categories beyond the annotated label space \cite{wu2024towards}. It has been widely used in various tasks and domains, including object detection \cite{zareian2021open, gu2021open}, segmentation \cite{ghiasi2022scaling, li2021language}, and scene understanding \cite{li2021language, ding2023pla}. For example, the object detection dataset COCO \cite{lin2014microsoft} contains only 80 categories. However, the objects in this world are almost infinite, further enhancing the importance of open vocabulary learning in this area. In this paper, we make the first attempt to address multimodal emotion recognition in an open-vocabulary manner. Compared with other tasks, multimodal emotion recognition is more challenging, where we need to consider multimodal and temporal information simultaneously. Meanwhile, open vocabulary learning is only one target of EMER. Considering the complexity of emotions, we also provide the evidence and reasoning process to improve the reliability of annotations.

\section{Dataset Construction}
\label{sec:3}
We build our dataset based on MER2023 \cite{lian2023mer}, a widely used corpus in multimodal emotion recognition. During the annotation process, we need to annotate multi-faceted clues, which requires a lot of manual effort. To reduce costs, we select 332 samples from MER2023 for annotation. In the future, we will explore ways to reduce costs and expand the dataset size. In this section, we introduce the data annotation process and analyze the multi-faceted capabilities of the annotated results.

\begin{figure}[t]
	\centering
	\includegraphics[width=0.9\linewidth]{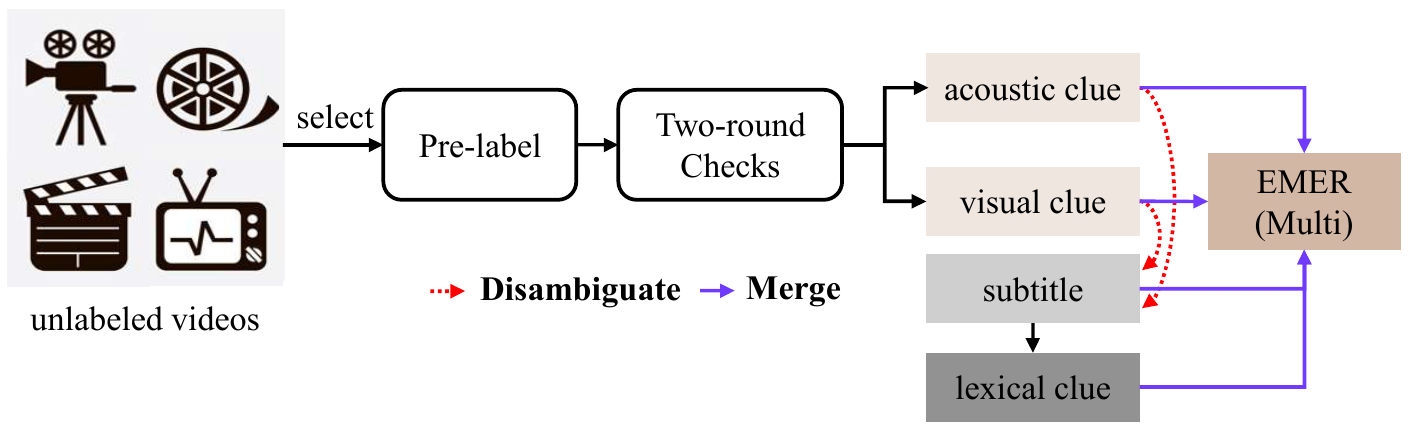}
	\caption{Pipeline of generating multimodal descriptions EMER(Multi).}
	\label{Figure2}
\end{figure}

\subsection{Data Annotation}
\label{sec:3-1}
We have some basic findings during the annotation process: \textcolor[rgb]{0.93,0.0,0.47}{\emph{Video subtitle is generally short, colloquial, and has vague emotional expressions. But by combining visual and acoustic clues, we can disambiguate the subtitle and generate more accurate descriptions.}} Therefore, we mainly annotate visual and acoustic clues and then use LLMs for disambiguation. Figure \ref{Figure2} shows the pipeline of data annotation and Table \ref{Table1} provides prompts involved in this process. Additionally, we provide an example to visualize the output of each step (see Appendix \ref{appendix:annotation}).

\paragraph{Pre-labeling.} Initially, we attempt to annotate visual and acoustic clues directly. However, the description obtained in this way is generally short and cannot cover all clues. Therefore, we use GPT-4V to generate initial annotations. Considering that GPT-4V does not support videos but only images, we sample the video and use the prompt (see \#1 in Table \ref{Table1}) to extract visual clues. To get acoustic clues, we try converting the audio to a mel-spectrogram, but GPT-4V fails to generate proper responses on the mel-spectrogram. Considering that the subtitle in audio also contains emotion-related clues, we use the prompt (see \#2 in Table \ref{Table1}), and its output is treated as the acoustic clues.

\paragraph{Two-round Checks.} During the proofreading process, we find some errors in the pre-labeled visual and acoustic clues. For visual clues, GPT-4V may produce hallucinatory responses, i.e., it may contain some clues that do not exist. For acoustic clues, the textual content is usually brief and colloquial. Without incorporating multimodal information, the clue merely based on the textual content may be incorrect. Additionally, there are repeated expressions and some key clues are missing. To obtain more reliable clues, we conduct two rounds of manual checks.

\paragraph{Disambiguation.} To obtain lexical clues, we use the checked acoustic and visual clues to disambiguate the subtitle (see Figure \ref{Figure2}). In this process, we rely on GPT-3.5 and use the \#3 prompt in Table \ref{Table1}. With its powerful reasoning ability, we can generate accurate lexical clues. Finally, we combine all clues and generate multimodal descriptions. These descriptions are noted as EMER(Multi).

\begin{table}[t]
	\centering
	\renewcommand\arraystretch{1.2}
	\caption{Prompts involved in data annotation.}
	\label{Table1}
	\scalebox{0.8}{
		\begin{tabular}{p{3.7cm}|p{12.8cm}}
			\hline
			Function & Prompt \\
			\hline
			
			\multirow{6}{*}{\#1 Pre-label Visual Clue} &  As an expert in the field of emotions, please focus on facial expressions, body language, environmental cues, and events in the video and predict the emotional state of the character. Please ignore the character's identity. We uniformly sample 3 frames from this video. Please consider the temporal relationship between these frames and provide a complete description of this video. Avoid using descriptions like ``the first image'' and ``the second image'', and instead use terms like ``beginning'', ``middle'', and ``end'' to denote the progression of time. \\
			
			\hline
			
			\multirow{3}{*}{\#2 Pre-label Acoustic Clue} & Please assume the role of an expert in the field of emotions. We have a piece of text. Please analyze which parts of it can be used to infer the emotional states of the characters, and provide reasoning for your inference. \\
			
			\hline
			
			\multirow{5}{*}{\#3 Disambiguate} & Please assume the role of an expert in the field of emotions. We provide audio and video cues that may be related to the emotions of the characters. Additionally, we provide the original subtitle of the video. Please analyze which parts of the subtitle can be used to infer the emotional states of the characters and provide reasoning for your inference. In the process of inference, please integrate the audio and video cues for analysis. \\
			
			\hline
		\end{tabular}
	}
\end{table}

\subsection{Annotation Analysis}
EMER(Multi) contains multi-modal emotion-related clues. From them, we can extract multi-faceted results, including visual clues, discrete emotions, valence scores, and open-vocabulary emotion labels. To realize these functions, we rely on GPT-3.5 and use the prompts in Table \ref{Table2}.

\paragraph{Visual Clue Analysis.} 
EMER(Multi) contains a variety of visual clues. In this section, we provide a statistical analysis of the number of visual clues. To extract visual clues, we use the prompt \#1 in Table \ref{Table2}. Experimental results demonstrate that each sample has an average of 4.95 visual clues, suggesting that EMER(Multi) contains rich clues for emotion recognition.

\paragraph{Discrete Emotion Recognition.}
Then, we attempt to reveal whether discrete emotions can be identified from EMER(Multi). Considering that our dataset is based on MER2023, which provides relatively accurate discrete labels, we treat its label as the ground truth. To identify emotions from EMER(Multi), we use the \#2 prompt in Table \ref{Table2} and restrict the label space to be consistent with MER2023. Experimental results show that the Top-1 and Top-2 accuracy can reach 93.48 and 96.89, respectively. Through further analysis, these errors are mainly caused by inaccurate labels in MER2023 or ranking errors of GPT-3.5. Therefore, we can conclude that EMER(Multi) contains clues for discrete emotion recognition.

\paragraph{Valence Estimation.} 
In addition to discrete emotion recognition, we also validate the valence estimation results based on EMER(Multi). Considering that MER2023 provides relatively accurate valence scores, we treat its label as the ground truth. To estimate the valence from EMER(Multi), we use the \#3 prompt in Table \ref{Table2} and set the score range -5$\sim$5, consistent with the range in MER2023. Then, we calculate the PCC value between MER2023-based and EMER-based valence scores. Experimental results show that their PCC can reach 0.88, a relatively high level, indicating that EMER(Multi) also contains clues for valence estimation.

\paragraph{OV Emotion Recognition.}
One of the main purposes behind EMER is to obtain richer emotions. Therefore, we extract all emotion labels from EMER(Multi) using the \#4 prompt in Table \ref{Table2}. In this process, we do not restrict the label space and predict emotions in an open-vocabulary manner. We perform a statistical analysis on the number of extracted labels. There are a total of 301 candidate labels, far more than 6 candidate labels in MER2023. Meanwhile, each sample has an average of 3 labels. Therefore, EMER(Multi) contains richer emotion labels than previous datasets.

In summary, EMER can unify two types of tasks: discrete emotion recognition and valence estimation. Compared with traditional emotion recognition, EMER can obtain richer emotion labels in an open-vocabulary manner. Meanwhile, it contains various clues that help determine emotional states.

\begin{table}[t]
	\centering
	\renewcommand\arraystretch{1.2}
	\caption{Prompts involved in annotation analysis.}
	\label{Table2}
	\scalebox{0.8}{
		\begin{tabular}{p{3.46cm}|p{13.1cm}}
			\hline
			Function & Prompt \\
			
			\hline
			
			\multirow{3}{*}{\#1 Visual Clue Analysis} & Please assume the role of an expert in the field of emotions. We provide clues related to the emotions of the characters in the video. Please output the facial movements and body gestures involved in the description, separated by commas. The output format should be in list form. \\
			
			\hline
			
			\multirow{4}{*}{\#2 Discrete Emotion Rec.} &  Please assume the role of an expert in the emotional domain. We provide clues that may be related to the emotions of the character. Based on the provided clues, identify the emotional states of the main characters. We provide a set of emotional candidates, please rank them in order of likelihood from high to low. The candidate set is \{Candidate Labels\}. \\
			
			\hline
			
			\multirow{8}{*}{\#3 Valence Estimation} &  As an expert in the emotional domain, we provide clues that may be related to the emotions of characters. Based on the provided clues, please identify the overall positive or negative emotional polarity of the main characters. The output should be a floating-point number ranging from -5 to +5. Here, -5 indicates extremely negative emotions, 0 indicates neutral emotions, and +5 indicates extremely positive emotions. Larger numbers indicate more positive emotions, while smaller numbers indicate more negative emotions. Please provide your judgment as a floating-point number with two decimal places, directly outputting the numerical result without including the analysis process. \\
			
			\hline
			
			\multirow{5}{*}{\#4 OV Emotion Rec.} & Please assume the role of an expert in the field of emotions. We provide clues that may be related to the emotions of the characters. Based on the provided clues, please identify the emotional states of the main characters. Please separate different emotional categories with commas and output only the clearly identifiable emotional categories in a list format. If none are identified, please output an empty list. \\
			
			\hline
		\end{tabular}
	}
\end{table}

\section{Experimental Setup}
\label{sec:4}

\subsection{Baselines}
\label{sec:4-1}
Considering that MLLMs can address various multimodal tasks, we attempt to use them to solve EMER. Since emotion recognition relies on temporal information, we choose MLLMs that support at least video or audio. Appendix \ref{appendix_sec:mllm} provides model cards of MLLMs involved in this paper. To build MLLMs, a mainstream idea is to align pre-trained models of other modalities to LLMs. For example, VideoChat \cite{li2023videochat} uses Q-Former \cite{li2023blip} to map visual queries into textual embedding space. SALMONN \cite{tang2023salmonn} proposes a window-level Q-Former to align speech and audio encoders with LLMs. After instruction fine-tuning, MLLMs can understand instructions and multimodal inputs.

To generate EMER-like descriptions using MLLMs, we first use the prompt in Appendix \ref{appendix_sec:mllm} (the prompt without subtitles) and its output is denoted $C$. Considering that the subtitle contains important clues for emotion recognition, we follow the disambiguation process in Figure \ref{Figure2} and use the clue $C$ to disambiguate the subtitle. For a fair comparison, we use similar prompts for audio, video, and audio-video LLMs. In Section \ref{sec:5}, we further investigate the role of subtitles and discuss different ways to integrate them. Please refer to the corresponding section for more details.

\subsection{Evaluation Metrics}
\label{sec:4-2}
One of the main purposes of EMER is to identify richer emotion labels. In this paper, we use the overlap rate between the predicted and annotated label sets as the evaluation metric. In addition, we also calculate some matching-based metrics, including BLEU$_{1}$, BLEU$_{4}$, METEOR, and ROUGE$_l$.

\paragraph{Emotion Recognition.} 
Since we do not fix the label space, MLLMs may generate synonyms, i.e., labels with different expressions but similar meanings (such as \emph{happy} and \emph{joy}). These synonyms affect the overlap rate between the annotated and predicted label sets. To reduce their impact, we first use GPT-3.5 to group all labels before metric calculation:

\textcolor[rgb]{0.93,0.0,0.47}{\emph{Please assume the role of an expert in the field of emotions. We provide a set of emotions. Please group the emotions, with each group containing emotions with the same meaning. Directly output the results. The output format should be a list containing multiple lists.}}

After that, we get a function $G(\cdot)$ that can map each label to its group ID. Suppose $\{y_i\}_{i=1}^M$ and $\{\hat{y}_i\}_{i=1}^N$ are the annotated and predicted label sets respectively, where $M$ and $N$ are the number of labels. To reduce the impact of synonyms, we first map each label into its group ID: $\mathcal{Y} = \{G(x) |x \in \{y_i\}_{i=1}^M\}$ and $\hat{\mathcal{Y}} = \{G(x) |x \in \{\hat{y}_i\}_{i=1}^N\}$. Then, we define the following two metrics:
\begin{equation}
\mbox{Accuracy}_{\mbox{s}} = \frac{|\mathcal{Y} \cap \hat{\mathcal{Y}}|}{|\hat{\mathcal{Y}}|}, \;\mbox{Recall}_{\mbox{s}} = \frac{|\mathcal{Y} \cap \hat{\mathcal{Y}}|}{|\mathcal{Y}|}.
\end{equation}

These two metrics are similar to traditional precision and recall but are defined at the set level. $\mbox{Accuracy}_{\mbox{s}}$ denotes how many predicted labels are correct; $\mbox{Recall}_{\mbox{s}}$ indicates whether the predicted results cover all annotated labels. We use the average of these two metrics for the final ranking:
\begin{equation}
\mbox{Avg} = \frac{\mbox{Accuracy}_{\mbox{s}} + \mbox{Recall}_{\mbox{s}}}{2}.
\end{equation}

\paragraph{Word-level Matching.}
Besides metrics for emotion recognition, we also calculate some typical metrics for natural language generation, including BLEU$_{1}$, BLEU$_{4}$, METEOR, and ROUGE$_l$. The main purpose behind them is that emotion-based metrics require OpenAI API call costs. We try to analyze whether there is also a strong correlation between matching-based and emotion-based metrics. If so, we can only calculate matching-based metrics to reduce the evaluation costs.

\subsection{Implementation Details}
\label{sec:4-3}
This paper uses the closed-source GPT for dataset construction and metric calculation. In this process, GPT-3.5 \cite{openai2022chatgpt} (``gpt-3.5-turbo-16k-0613'') and GPT-4V \cite{openai2023gpt4v} (``gpt-4-vision-preview'') perform similarly in plain text analysis. To reduce API call costs, we use GPT-3.5 for text processing and GPT-4V for image processing. We run all experiments twice and report the average score and standard deviation. For baseline MLLMs, we use their original 7B weights. All models are implemented with PyTorch and all inference processes are carried out with a 32G NVIDIA Tesla V100 GPU.

\section{Results and Discussion}
\label{sec:5}
EMER aims to achieve reliable and accurate emotion recognition. This paper mainly focuses on accuracy and the reliability analysis is left to our future work. In this section, we first reveal the impact of language on the evaluation metrics. Then, we report the performance of MLLMs and conduct ablation studies around modality influence and how to integrate subtitles. Finally, we reveal the relationship between one-hot and OV labels and visualize the correlation between different metrics.

\paragraph{Language Influence.}
\begin{wrapfigure}{r}{0cm}
	\centering
	\includegraphics[width=0.46\textwidth]{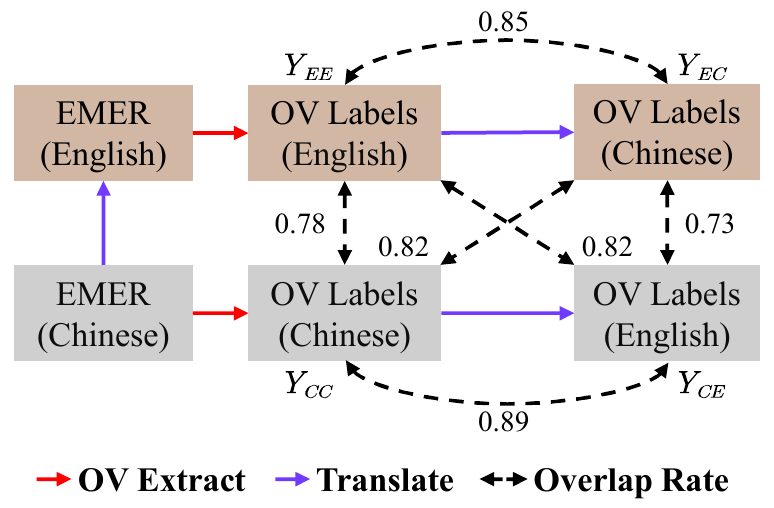}
	\caption{Language influence analysis.}
	\label{Figure3}
\end{wrapfigure}
The initial EMER dataset is in Chinese and we use GPT-3.5 to translate it into English. Therefore, our dataset has English and Chinese versions. In this section, we attempt to reveal the language influence on the evaluation metrics. As shown in Figure \ref{Figure3}, we first extract emotion labels $Y_{EE}$ and $Y_{CC}$ from the English and Chinese datasets, respectively. Then, we translate these labels into another language and get $Y_{EC}$ and $Y_{CE}$. Referring to Section \ref{sec:4-2}, we define a metric called \emph{overlap rate} to measure label similarity. Specifically, assume $\{p_i^1\}_{i=1}^{N_1}$ and $\{p_i^2\}_{i=1}^{N_2}$ are two label sets and $G(\cdot)$ is the synonym mapping function. We first map each label to its group ID: $\mathcal{P}^1 = \{G(x) |x \in \{p_i^1\}_{i=1}^{N_1}\}$ and $\mathcal{P}^2 = \{G(x) |x \in \{p_i^2\}_{i=1}^{N_2}\}$ and calculate the following metric:
\begin{equation}
\mbox{Overlap}_{\mbox{s}} = \frac{|\mathcal{P}^1 \cap \mathcal{P}^2|}{|\mathcal{P}^1 \cup \mathcal{P}^2|}.
\end{equation}
A higher overlap rate indicates a higher degree of label similarity. From Figure \ref{Figure3}, we observe some interesting phenomena: 1) Translation reduces the overlap rate. For example, $Y_{EE}$ to $Y_{EC}$ results in a 0.15 decrease in the overlap rate. The main reason lies in the increased difficulty of label grouping, i.e., $G(\cdot)$, in the cross-language setting. In some cases, we find that the grouping process may be based on language type rather than label similarity.

2) There are certain differences in labels extracted from descriptions in different languages. For example, under the same language setup, $Y_{EE}$ to $Y_{CE}$ results in a 0.18 decrease in the overlap rate. The reason may lie in some differences in the definition of emotion in distinct languages. \textbf{To obtain more accurate labels, we merge labels extracted from both languages and perform manual checks.} These checked labels are regarded as ground truth, noted as $Y_{gt}$.

3) If we extract labels from descriptions in different languages and calculate overlap rates in a cross-language setup, it will cause the largest drop. For example, $Y_{EE}$ to $Y_{CC}$ (or $Y_{EC}$ to $Y_{CE}$) results in a reduction of 0.22 (or 0.27). These results further confirm the two findings mentioned above.

\paragraph{Main Results.}
In this section, we report the emotion recognition results of different methods. Besides MLLMs, we introduce two heuristic baselines: \emph{Empty} and \emph{Random}. For the former, we predict each sample as ``unable to judge the emotional state''. For the latter, we randomly select a label from the MER2023's candidate set (i.e., \emph{worried}, \emph{happy}, \emph{neutral}, \emph{angry}, \emph{surprised}, \emph{sad}) and generate a description like ``through the video, we can judge the emotional state is \emph{\{emotion\}}''. These two baselines reflect performance lower bounds. According to our previous findings, there are certain differences in labels extracted from descriptions in distinct languages. Therefore, we report results in both Chinese and English. Specifically, take Figure \ref{Figure3} as an example. Under the Chinese condition, we calculate the metrics on the English version of $Y_{gt}$ and $Y_{EE}$; under the English condition, we calculate the metrics on the English version of $Y_{gt}$ and $Y_{CE}$.

Experimental results in Table \ref{Table3} demonstrate that MLLMs outperform the heuristic baselines, indicating that MLLMs can address emotion recognition to some extent. However, there is still a significant performance gap between the predictions of MLLMs and the ground truth $Y_{gt}$, which highlights the limitations of existing MLLMs and the difficulty of this task. Meanwhile, models that perform well in Chinese generally perform well in English. These results suggest that language differences have a limited impact on performance rankings.

\begin{table}[t]
	\centering
	\renewcommand\tabcolsep{4.6pt}
	\renewcommand\arraystretch{1.06}
	\caption{Main results on emotion recognition. We consider language influence and report results for descriptions in distinct languages. The values in the gray column are used for the final ranking.}
	\label{Table3}
	\scalebox{0.8}{
		\begin{tabular}{lccc|>{\columncolor{lightgray}}ccc|>{\columncolor{lightgray}}ccc}
			\hline
			\multirow{2}{*}{Model} & \multirow{2}{*}{L} & \multirow{2}{*}{V} & \multirow{2}{*}{A} & \multicolumn{3}{c|}{English} & \multicolumn{3}{c}{Chinese} \\
			&&&&Avg & $\mbox{Accuracy}_{\mbox{s}}$ & $\mbox{Recall}_{\mbox{s}}$ & Avg & $\mbox{Accuracy}_{\mbox{s}}$ & $\mbox{Recall}_{\mbox{s}}$\\
			\hline
			Empty          & $\times$ & $\times$ & $\times$&0.00$\pm$0.00 & 0.00$\pm$0.00 & 0.00$\pm$0.00 & 0.00$\pm$0.00 & 0.00$\pm$0.00 & 0.00$\pm$0.00 \\
			Random         & $\times$ & $\times$ & $\times$&19.13$\pm$0.06 & 24.85$\pm$0.15 & 13.42$\pm$0.04 & 18.59$\pm$0.00 & 24.70$\pm$0.00 & 12.48$\pm$0.00 \\
			\hline
			Qwen-Audio \cite{chu2023qwen}   & $\surd$  & $\times$ & $\surd$&40.23$\pm$0.09 & 49.42$\pm$0.18 & 31.04$\pm$0.00 & 43.53$\pm$0.04 & 53.71$\pm$0.00 & 33.34$\pm$0.09 \\
			OneLLM \cite{han2023onellm}     & $\surd$  & $\times$ & $\surd$&43.04$\pm$0.06 & 45.92$\pm$0.05 & 40.15$\pm$0.06 & 46.77$\pm$0.01 & 52.07$\pm$0.06 & 41.47$\pm$0.08 \\
			Otter  \cite{li2023otter}   & $\surd$  & $\surd$ & $\times$&44.40$\pm$0.09 & 50.71$\pm$0.10 & 38.09$\pm$0.09 & 46.92$\pm$0.04 & 52.65$\pm$0.16 & 41.18$\pm$0.08 \\
			VideoChat  \cite{li2023videochat} & $\surd$  & $\surd$ & $\times$&45.70$\pm$0.09 & 42.90$\pm$0.27 & 48.49$\pm$0.10 & 45.63$\pm$0.04 & 47.20$\pm$0.12 & 44.05$\pm$0.05 \\
			Video-LLaMA \cite{zhang2023video}  & $\surd$ & $\surd$ & $\times$&44.74$\pm$0.14 & 44.14$\pm$0.13 & 45.34$\pm$0.15 & 47.27$\pm$0.03 & 47.98$\pm$0.07 & 46.56$\pm$0.01 \\
			PandaGPT  \cite{su2023pandagpt}     & $\surd$  & $\surd$  & $\surd$&46.21$\pm$0.17 & 50.03$\pm$0.01 & 42.38$\pm$0.33 & 47.88$\pm$0.02 & 53.01$\pm$0.08 & 42.75$\pm$0.11 \\
			SALMONN \cite{tang2023salmonn}      & $\surd$  & $\times$ & $\surd$&48.06$\pm$0.04 & 50.20$\pm$0.04 & 45.92$\pm$0.04 & 48.53$\pm$0.03 & 52.24$\pm$0.00 & 44.82$\pm$0.05 \\
			Video-LLaVA \cite{lin2023video}   & $\surd$  & $\surd$ & $\times$&47.12$\pm$0.15 & 48.58$\pm$0.02 & 45.66$\pm$0.29 & 49.59$\pm$0.05 & 53.95$\pm$0.03 & 45.23$\pm$0.13 \\
			VideoChat2 \cite{li2024mvbench}   & $\surd$  & $\surd$ & $\times$&49.60$\pm$0.28 & 54.72$\pm$0.41 & 44.47$\pm$0.15 & 49.90$\pm$0.06 & 57.12$\pm$0.08 & 42.68$\pm$0.04 \\
			OneLLM \cite{han2023onellm}     & $\surd$  & $\surd$  & $\times$&50.99$\pm$0.08 & 55.93$\pm$0.09 & 46.06$\pm$0.06 & 51.84$\pm$0.08 & 56.43$\pm$0.04 & 47.26$\pm$0.11 \\
			LLaMA-VID \cite{li2023llama}    & $\surd$  & $\surd$  & $\times$&51.29$\pm$0.09 & 52.71$\pm$0.18 & 49.87$\pm$0.00 & 52.45$\pm$0.02 & 57.30$\pm$0.00 & 47.61$\pm$0.03 \\
			mPLUG-Owl \cite{ye2023mplug}   & $\surd$  & $\surd$ & $\times$&52.79$\pm$0.13 & 54.54$\pm$0.13 & 51.04$\pm$0.13 & 51.43$\pm$0.03 & 56.40$\pm$0.11 & 46.47$\pm$0.18 \\
			Video-ChatGPT \cite{maaz2023video} & $\surd$ & $\surd$ & $\times$&50.73$\pm$0.06 & 54.03$\pm$0.04 & 47.44$\pm$0.07 & 55.34$\pm$0.02 & 61.15$\pm$0.10 & 49.52$\pm$0.06 \\
			Chat-UniVi \cite{jin2023chat}   & $\surd$  & $\surd$  & $\times$&53.09$\pm$0.01 & 53.68$\pm$0.00 & 52.50$\pm$0.02 & 54.20$\pm$0.02 & 58.54$\pm$0.01 & 49.86$\pm$0.03 \\
			GPT-4V \cite{openai2023gpt4v}     & $\surd$   & $\surd$ & $\times$&56.69$\pm$0.04 & 48.52$\pm$0.07 & 64.86$\pm$0.00 & 57.34$\pm$0.01 & 54.61$\pm$0.02 & 60.07$\pm$0.01 \\
			\hline
			EMER(Text)  & $\surd$  & $\times$ & $\times$&47.13$\pm$0.08 & 54.41$\pm$0.15 & 39.84$\pm$0.01 & 44.09$\pm$0.24 & 50.69$\pm$0.26 & 37.50$\pm$0.23 \\
			EMER(Video) & $\times$ & $\surd$  & $\times$&60.67$\pm$0.12 & 63.29$\pm$0.08 & 58.05$\pm$0.16 & 62.05$\pm$0.10 & 66.47$\pm$0.13 & 57.62$\pm$0.08 \\
			EMER(Audio) & $\surd$  & $\times$ & $\surd$&65.42$\pm$0.04 & 67.54$\pm$0.08 & 63.30$\pm$0.00 & 68.59$\pm$0.07 & 70.10$\pm$0.06 & 67.07$\pm$0.08 \\
			EMER(Multi) & $\surd$  & $\surd$  & $\surd$&80.05$\pm$0.24 & 80.03$\pm$0.37 & 80.07$\pm$0.10 & 85.20$\pm$0.03 & 87.09$\pm$0.00 & 83.31$\pm$0.05 \\
			\hline
		\end{tabular}
	}
\end{table}

\begin{figure*}[!t]
	\centering
	\includegraphics[width=\linewidth]{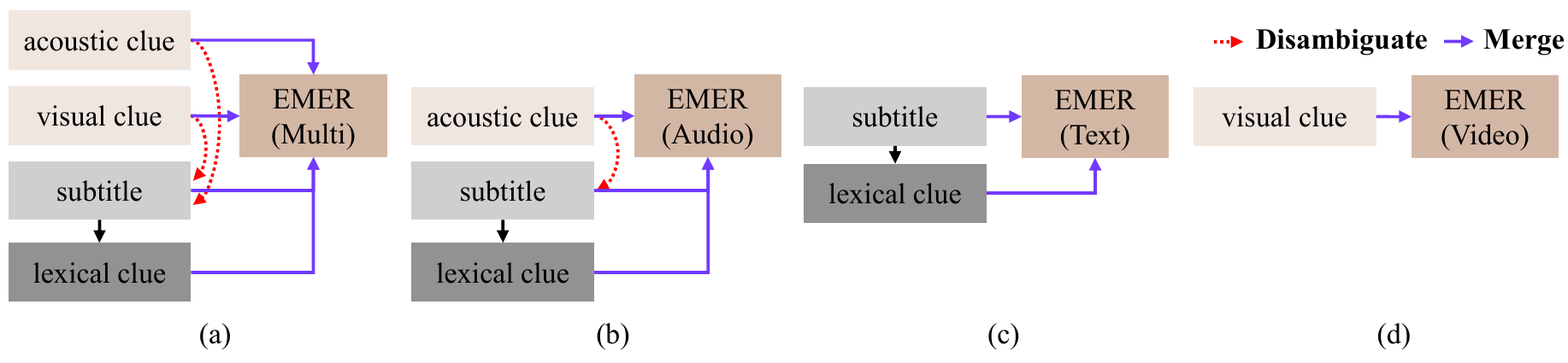}
	\caption{Pipeline for generating unimodal and multimodal descriptions.}
	\label{Figure4}
\end{figure*}

\paragraph{Impact of Modality.}
EMER(Multi) uses visual and acoustic clues to disambiguate subtitles and generate lexical clues. To study the impact of modality, we further generate three descriptions: EMER(Audio), EMER(Text), and EMER(Video). The generation process is shown in Figure \ref{Figure4}. Specifically, for EMER(Audio), we only use the acoustic clue to disambiguate subtitles. For EMER(Text), we infer the emotional state from subtitles and use the \#2 prompt in Table \ref{Table1} to generate lexical clues. Meanwhile, we directly use the visual clue as EMER(Video). 

In Table \ref{Table3}, we observe that EMER(Multi) can achieve the best performance in emotion recognition. The reason lies in that emotions are conveyed through various modalities. Combining all clues can realize more accurate emotion recognition. Meanwhile, EMER(Text) performs worst among the four descriptions. This also validates our basic principle in dataset construction (see Section \ref{sec:3-1}). That is, the subtitle is relatively vague. However, by incorporating clues from other modalities, we can disambiguate the subtitle and generate more accurate lexical clues. Furthermore, we observe that EMER(Audio) performs better than EMER(Video). The reason lies in the samples in our dataset focusing more on audio to convey emotions, which is consistent with previous findings \cite{lian2024merbench}.

\begin{figure*}[t]
	\begin{center}
		\subfigure[Otter]{
			\label{Figure5-1}
			\centering
			\includegraphics[width=0.23\linewidth, trim=0 0 0 0]{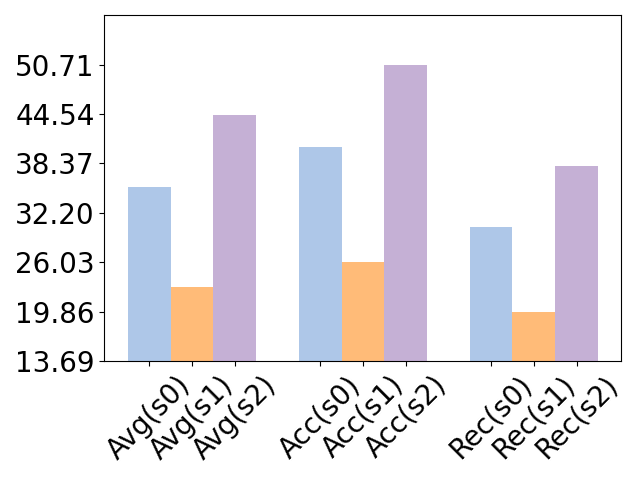}
		} 
		\subfigure[PandaGPT]{
			\label{Figure5-2}
			\centering
			\includegraphics[width=0.23\linewidth, trim=0 0 0 0]{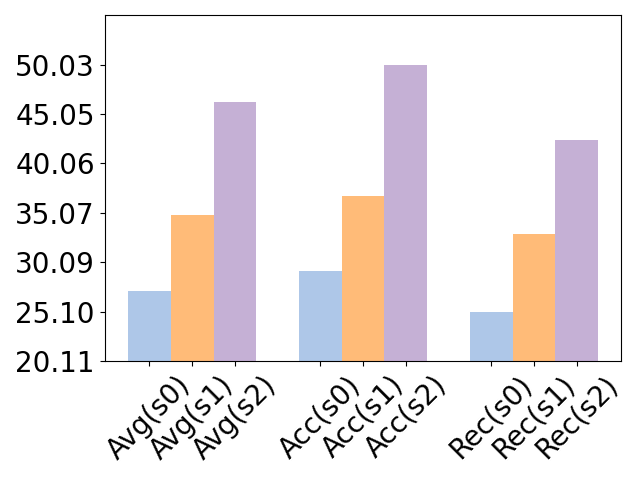}
		} 
		\subfigure[Video-ChatGPT]{
			\label{Figure5-3}
			\centering
			\includegraphics[width=0.23\linewidth, trim=0 0 0 0]{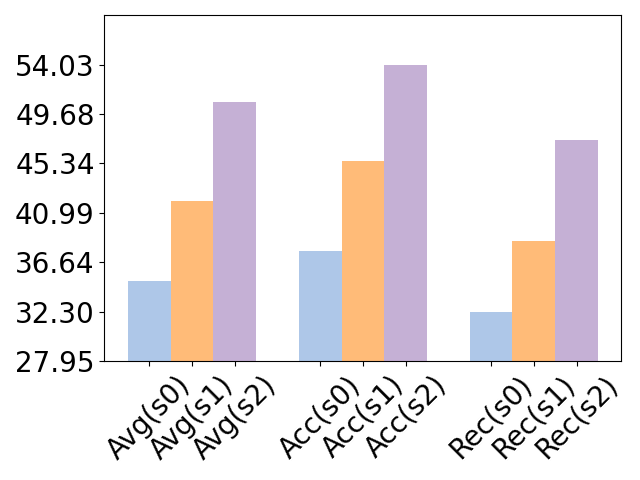}
		}  
		\subfigure[Video-LLaMA]{
			\label{Figure5-4}
			\centering
			\includegraphics[width=0.23\linewidth, trim=0 0 0 0]{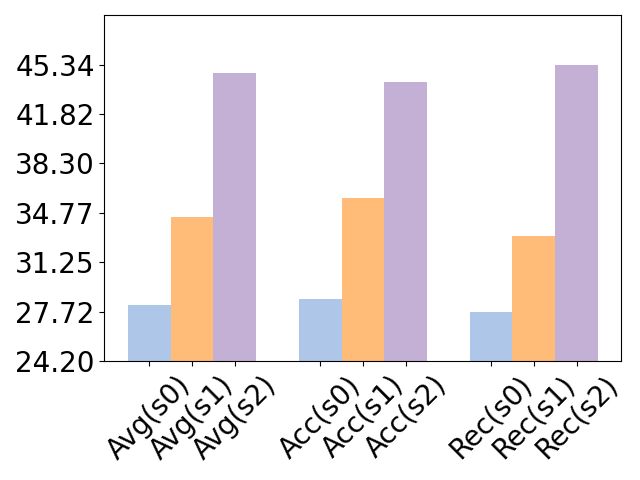}
		} 
		\subfigure[VideoChat]{
			\label{Figure5-5}
			\centering
			\includegraphics[width=0.23\linewidth, trim=0 0 0 0]{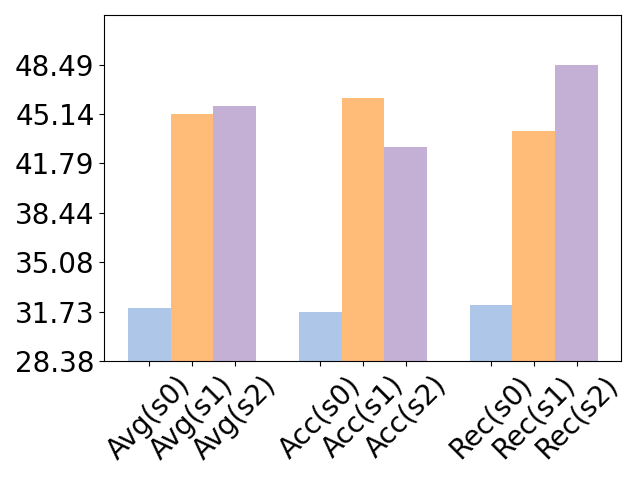}
		}  
		\subfigure[VideoChat2]{
			\label{Figure5-6}
			\centering
			\includegraphics[width=0.23\linewidth, trim=0 0 0 0]{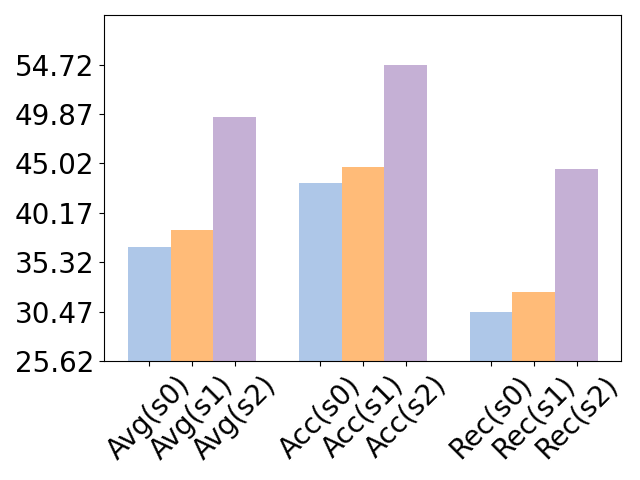}
		} 
		\subfigure[mPLUG-Owl]{
			\label{Figure5-7}
			\centering
			\includegraphics[width=0.23\linewidth, trim=0 0 0 0]{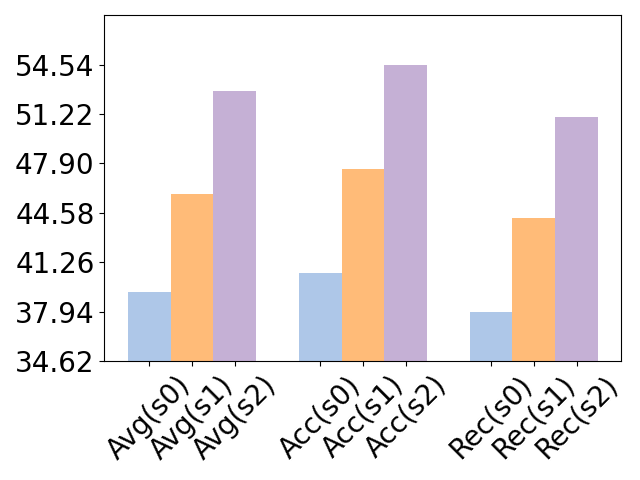}
		}  	
		\subfigure[SALMONN]{
			\label{Figure5-8}
			\centering
			\includegraphics[width=0.23\linewidth, trim=0 0 0 0]{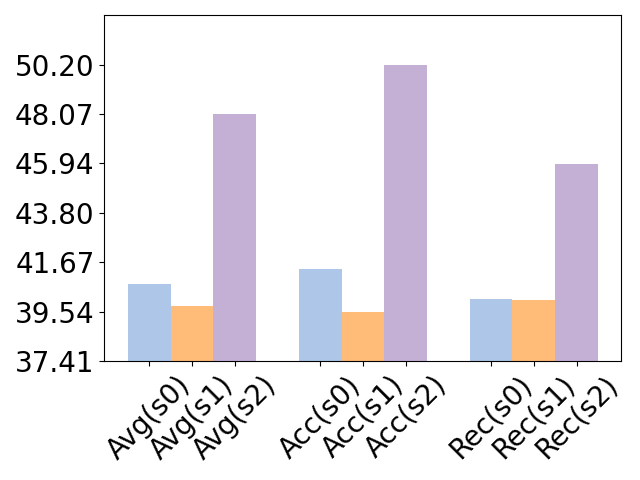}
		}  
		\subfigure[Qwen-Audio]{
			\label{Figure5-9}
			\centering
			\includegraphics[width=0.23\linewidth, trim=0 0 0 0]{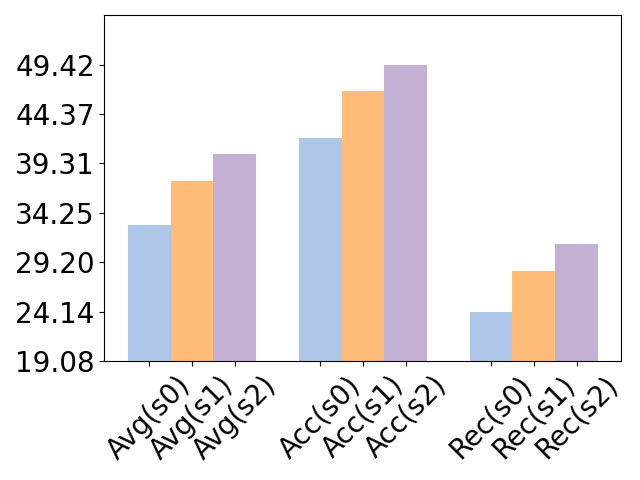}
		}  
		\subfigure[Video-LLaVA]{
			\label{Figure5-10}
			\centering
			\includegraphics[width=0.23\linewidth, trim=0 0 0 0]{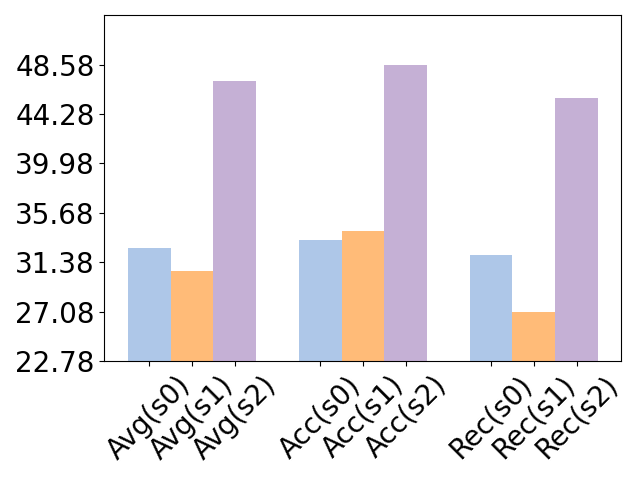}
		}  
		\subfigure[LLaMA-VID]{
			\label{Figure5-11}
			\centering
			\includegraphics[width=0.23\linewidth, trim=0 0 0 0]{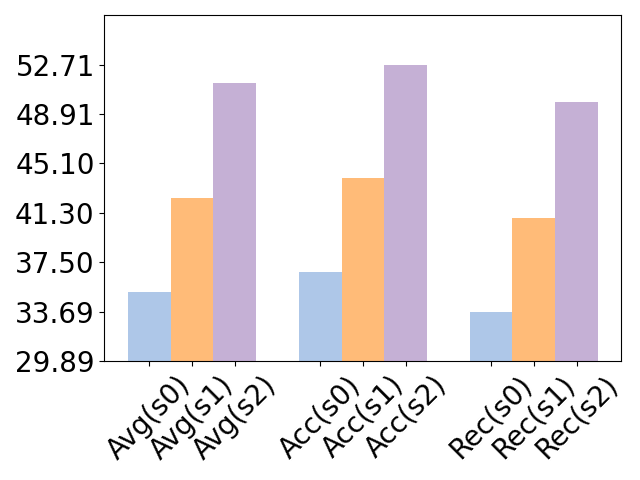}
		} 
		\subfigure[Chat-UniVi]{
			\label{Figure5-12}
			\centering
			\includegraphics[width=0.23\linewidth, trim=0 0 0 0]{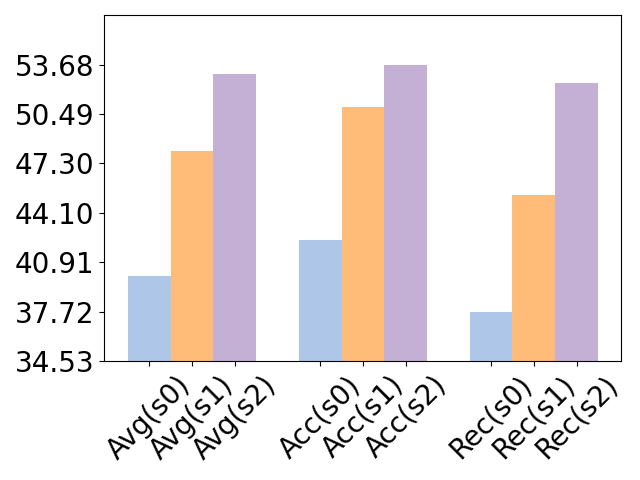}
		}		
	\end{center}
	\caption{Performance of different subtitle integration strategies on varying MLLMs.}
	\label{Figure5}
\end{figure*}

\paragraph{Impact of Subtitles.}
To generate EMER-like descriptions using MLLMs, this paper uses the prompts without subtitles to extract clues and then exploits the extracted clues to disambiguate subtitles, noted as ``S2''. In this section, we reveal the impact of different ways to integrate subtitles and introduce two additional baselines: 1) S0 uses the prompts without subtitles to generate descriptions; 2) S1 uses the prompts with subtitles to generate descriptions. More details about these prompts can be found in Appendix \ref{appendix_sec:mllm}. Specifically, S2 is equivalent to first using S0 to extract clues and then using these clues to disambiguate subtitles. This disambiguation process relies on GPT-3.5. Compared with S2, S1 merges two steps in one, taking into account subtitles and other clues simultaneously.

Figure \ref{Figure5} shows the emotion recognition results of different strategies. More results can be found in Appendix \ref{appendix_sec:subtitle}. From these figures, we observe that S1 and S2 generally perform better than S0. These results demonstrate the importance of subtitles in emotion recognition. Meanwhile, S2 generally outperforms S1. The reason lies in that including subtitles in the prompt makes the prompt more complex. However, current open-source MLLMs may have difficulty understanding complex prompts, resulting in limited performance. By separating this process into two steps, S2 can reduce the task difficulty and achieve better performance. These results also demonstrate the rationality of this paper using S2 as the default strategy to integrate subtitles.

\paragraph{One-hot vs. OV Labels.}
This section reveals the relationship between MER2023-based one-hot labels and OV labels $Y_{gt}$. In Table \ref{Table4}, we observe that one-hot labels have relatively high \emph{accuracy} but relatively low \emph{recall}. These results show that the one-hot labels provided by MER2023 are generally correct. However, they cannot cover all emotions due to the limited label space and the limited number of labels. Meanwhile, these results demonstrate the necessity of using \emph{avg} for the final ranking, which ensures the generation of more accurate and comprehensive labels.

\begin{table}[h]
	\centering
	\caption{Performance of one-hot labels in OV emotion recognition.}
	\label{Table4}
	\scalebox{0.8}{
		\begin{tabular}{c|>{\columncolor{lightgray}}ccc}
			\hline
			Language & Avg & $\mbox{Accuracy}_{\mbox{s}}$ & $\mbox{Recall}_{\mbox{s}}$ \\
			\hline
			English & 71.61$\pm$0.04 & 92.17$\pm$0.00 & 51.05$\pm$0.08  \\
			Chinese & 72.20$\pm$0.01 & 93.07$\pm$0.00 & 51.32$\pm$0.03 \\
			\hline
		\end{tabular}
	}
\end{table}

\paragraph{Metric Correlation Analysis.}
\begin{wrapfigure}{r}{0cm}
	\centering
	\includegraphics[width=0.4\textwidth]{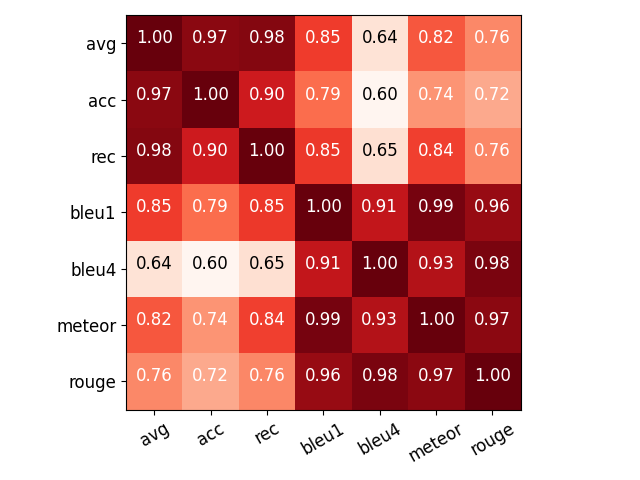}
	\caption{Metric correlation analysis.}
	\label{Figure6}
\end{wrapfigure}
In Table \ref{Table3}, we observe that EMER(Multi) achieves the best performance in emotion recognition. Therefore, a conjecture arises: whether descriptions ``closer'' to EMER(Multi) lead to better emotion recognition performance. The most common ways to measure the ``closeness'' between two sentences are matching-based metrics, such as BLEU$_{1}$, BLEU$_{4}$, METEOR, and ROUGE$_l$. Therefore, in this section, we reveal the correlation between two types of metrics: emotion-based and matching-based metrics. 

Figure \ref{Figure6} shows the PCC scores between different metrics. In this figure, we report the results in English. In Appendix \ref{appendix_sec:metric}, we provide results of other languages, as well as raw scores for matching-based metrics. From these results, we observe relatively high correlations within emotion-based (or matching-based) metrics. However, the correlation between these metrics is relatively low. Therefore, there are certain differences between them.

\section{Limitations and Societal Impacts}
\label{sec:6}
This paper proposes a new task EMER. Due to the high annotation cost, our initial dataset contains 332 samples. In the future, we will explore ways to reduce the cost and expand the dataset size. Meanwhile, we evaluate some typical MLLMs but do not cover all models. In the future, we will expand the evaluation scope. In addition, EMER aims to achieve reliable and accurate emotion recognition. This paper mainly focuses on accuracy. In the future, we will define more metrics and evaluate different MLLMs from the reliability perspective. Moreover, this paper focuses on the problem definition, dataset construction, metric definition, and evaluation. In the future, we will design more effective frameworks to solve this challenging task.

Emotion recognition technology has a positive impact on the development of human-computer interaction. However, machines that are too human-like may affect social stability and cause panic. Therefore, we need to monitor the development of this technology, although the current systems are still some distance away from fully human-like systems.

\section{Conclusion}
\label{sec:7}
This paper introduces a new task, EMER. Unlike traditional emotion recognition, EMER requires further evidence to support the prediction results. By introducing this task, we aim to improve the reliability and accuracy of emotion recognition technology. To facilitate further research, we construct an initial dataset, develop baselines, and define evaluation metrics. Then, we use MLLMs as the baselines to solve this task and evaluate their performance. Experimental results demonstrate that MLLMs struggle to achieve satisfactory results, indicating the difficulty of this task. Meanwhile, we systematically analyze the impact of language, modality, and subtitle integration strategy. We also reveal the correlations between different metrics. In summary, EMER contains multi-faceted multi-modal clues and can serve as a general format for emotion-related tasks.

\begin{ack}
This work is supported by the National Natural Science Foundation of China (NSFC) (No.62201572, No.62276259, No.U21B2010, No.62271083, No.62306316, No.62322120).
\end{ack}

\bibliography{mybib}
\bibliographystyle{unsrt}

\clearpage
\appendix

\section{Example for Data Annotation}
\label{appendix:annotation}
Our data annotation process involves three key steps: pre-labeling, two-round checks, and disambiguation. In this section, we provide an example to visualize the output of each step. From the generated EMER description, we can extract richer emotions in an open-vocabulary manner.

\begin{figure*}[h]
	\centering
	\includegraphics[width=\linewidth]{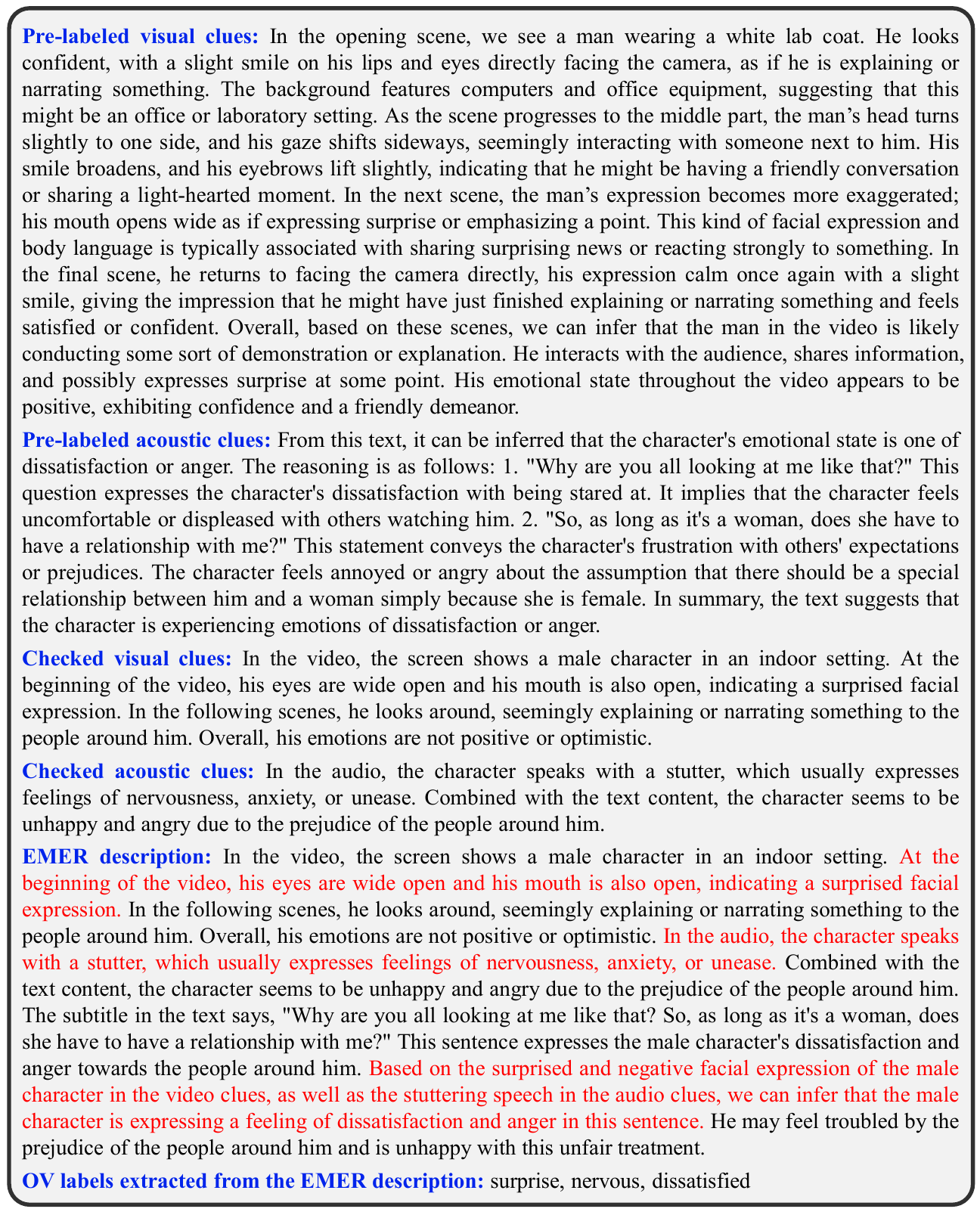}
	\caption{One example (``sample\_00000669'') to visualize the data annotation process.}
	\label{Figure7}
\end{figure*}

\clearpage

\section{Details about MLLMs}
\label{appendix_sec:mllm}
In this section, we provide model cards (see Table \ref{Table5}) and prompts (see Table \ref{Table6}) for MLLMs. For each LLM, we provide two prompts: one that ignores subtitles and one that considers subtitles.

\begin{table*}[h]
	\centering
	\renewcommand\arraystretch{1.06}
	\caption{Model cards for MLLMs.}
	\label{Table5}
	\scalebox{0.8}{
		\begin{tabular}{l|l|l}
			\hline
			Models & Support Modality &Link \\
			\hline
			Otter & Video, Text & \textcolor[rgb]{0.93,0.0,0.47}{https://github.com/Luodian/Otter} \\
		
			VideoChat & Video, Text & \textcolor[rgb]{0.93,0.0,0.47}{https://github.com/OpenGVLab/Ask-Anything/tree/main/video\_chat} \\
			VideoChat2 & Video, Text & \textcolor[rgb]{0.93,0.0,0.47}{https://github.com/OpenGVLab/Ask-Anything/tree/main/video\_chat2} \\
			Video-LLaVA & Video, Text & \textcolor[rgb]{0.93,0.0,0.47}{https://github.com/PKU-YuanGroup/Video-LLaVA} \\
			Video-LLaMA & Video, Text & \textcolor[rgb]{0.93,0.0,0.47}{https://github.com/DAMO-NLP-SG/Video-LLaMA} \\
			Video-ChatGPT & Video, Text & \textcolor[rgb]{0.93,0.0,0.47}{https://github.com/mbzuai-oryx/Video-ChatGPT} \\
			LLaMA-VID & Video, Text & \textcolor[rgb]{0.93,0.0,0.47}{https://github.com/dvlab-research/LLaMA-VID} \\
			mPLUG-Owl & Video, Text & \textcolor[rgb]{0.93,0.0,0.47}{https://github.com/X-PLUG/mPLUG-Owl} \\
			Chat-UniVi & Video, Text & \textcolor[rgb]{0.93,0.0,0.47}{https://github.com/PKU-YuanGroup/Chat-UniVi} \\
			\hline
			SALMONN & Audio, Text & \textcolor[rgb]{0.93,0.0,0.47}{https://github.com/bytedance/SALMONN} \\
			Qwen-Audio & Audio, Text & \textcolor[rgb]{0.93,0.0,0.47}{https://github.com/QwenLM/Qwen-Audio} \\
			\hline
			OneLLM & Audio, Video, Text & \textcolor[rgb]{0.93,0.0,0.47}{https://github.com/csuhan/OneLLM} \\
			PandaGPT & Audio, Video, Text & \textcolor[rgb]{0.93,0.0,0.47}{https://github.com/yxuansu/PandaGPT} \\
			\hline
			
		\end{tabular}
	}
\end{table*}

\begin{table}[h]
	\centering
	\renewcommand\arraystretch{1.2}
	\caption{Prompts for generating EMER-like descriptions using MLLMs. We provide two prompts: one that ignores subtitles and one that considers subtitles.}
	\label{Table6}
	\scalebox{0.8}{
		\begin{tabular}{p{2.6cm}|p{1cm}<{\centering}|p{12.2cm}}
			\hline
			Models & Subtitle & Prompt \\
			\hline
			
			\multirow{7}{*}{Audio LLM} & \multirow{3}{*}{$\times$} & As an expert in the field of emotions, please focus on the \textcolor[rgb]{0.93,0.0,0.47}{acoustic information} in the audio to discern clues related to the emotions of the individual. Please provide a detailed description and ultimately predict the emotional state of the individual. \\
			
			\cline{2-3}
			
			& \multirow{4}{*}{$\surd$} & \textcolor[rgb]{0.93,0.0,0.47}{Subtitle content of the audio: \{subtitle\};} As an expert in the field of emotions, please focus on the \textcolor[rgb]{0.93,0.0,0.47}{acoustic information and subtitle content} in the audio to discern clues related to the emotions of the individual. Please provide a detailed description and ultimately predict the emotional state of the individual in the audio.  \\
			 
			\hline
			
			\multirow{8}{*}{Video LLM} & \multirow{4}{*}{$\times$} & As an expert in the field of emotions, please focus on the \textcolor[rgb]{0.93,0.0,0.47}{facial expressions, body movements, environment, etc.,} in the video to discern clues related to the emotions of the individual. Please provide a detailed description and ultimately predict the emotional state of the individual in the video. \\
			
			\cline{2-3}
			
			& \multirow{4}{*}{$\surd$} & \textcolor[rgb]{0.93,0.0,0.47}{Subtitle content of the video: \{subtitle\};} As an expert in the field of emotions, please focus on the \textcolor[rgb]{0.93,0.0,0.47}{facial expressions, body movements, environment, subtitle content, etc.,} in the video to discern clues related to the emotions of the individual. Please provide a detailed description and ultimately predict the emotional state of the individual. \\
			
			\hline
			
			\multirow{9}{*}{Audio-Video LLM} & \multirow{4}{*}{$\times$} & As an expert in the field of emotions, please focus on the \textcolor[rgb]{0.93,0.0,0.47}{facial expressions, body movements, environment, acoustic information, etc.,} in the video to discern clues related to the emotions of the individual. Please provide a detailed description and ultimately predict the emotional state of the individual in the video. \\
			
			\cline{2-3}
			
			& \multirow{5}{*}{$\surd$} & \textcolor[rgb]{0.93,0.0,0.47}{Subtitle content of the video: \{subtitle\};} As an expert in the field of emotions, please focus on the \textcolor[rgb]{0.93,0.0,0.47}{facial expressions, body movements, environment, acoustic information, subtitle content, etc.,} in the video to discern clues related to the emotions of the individual. Please provide a detailed description and ultimately predict the emotional state of the individual in the video. \\
			
			\hline
		\end{tabular}
	}
\end{table}

\clearpage
\section{Impact of Subtitles}
\label{appendix_sec:subtitle}
In Table \ref{Table7}, we compare the emotion recognition results of different subtitle integration strategies. This table involves three strategies: S0, S1, and S2. Generally, S2 outperforms S0 and S1.

\begin{table}[h]
	\centering
	\renewcommand\tabcolsep{5pt}
	\renewcommand\arraystretch{1.06}
	\caption{Performance of different subtitle integration strategies.}
	\label{Table7}
	\scalebox{0.8}{
		\begin{tabular}{lc|ccc|ccc}
			\hline
			\multirow{2}{*}{Model} & \multirow{2}{*}{Strategy} & \multicolumn{3}{c|}{English} & \multicolumn{3}{c}{Chinese} \\
			&&Avg & $\mbox{Accuracy}_{\mbox{s}}$ & $\mbox{Recall}_{\mbox{s}}$ & Avg & $\mbox{Accuracy}_{\mbox{s}}$ & $\mbox{Recall}_{\mbox{s}}$\\
			\hline
			Otter \cite{li2023otter} & S0 & 35.45$\pm$0.02 & 40.41$\pm$0.03 & 30.48$\pm$0.01 & 31.61$\pm$0.11 & 35.71$\pm$0.15 & 27.51$\pm$0.07  \\
			Otter \cite{li2023otter} & S1 & 22.95$\pm$0.06 & 26.05$\pm$0.08 & 19.86$\pm$0.04 & 25.56$\pm$0.04 & 29.14$\pm$0.03 & 21.99$\pm$0.05  \\
			\rowcolor{lightgray}
			Otter \cite{li2023otter} & S2 & 44.40$\pm$0.09 & 50.71$\pm$0.10 & 38.09$\pm$0.09 & 46.92$\pm$0.04 & 52.65$\pm$0.16 & 41.18$\pm$0.08  \\
			\hline
			PandaGPT \cite{su2023pandagpt} & S0 & 27.14$\pm$0.02 & 29.18$\pm$0.08 & 25.10$\pm$0.04 & 28.85$\pm$0.01 & 30.95$\pm$0.00 & 26.76$\pm$0.03  \\
			PandaGPT \cite{su2023pandagpt} & S1 & 34.86$\pm$0.22 & 36.77$\pm$0.30 & 32.94$\pm$0.14 & 34.90$\pm$0.16 & 37.27$\pm$0.15 & 32.53$\pm$0.18  \\
			\rowcolor{lightgray}
			PandaGPT \cite{su2023pandagpt} & S2 & 46.21$\pm$0.17 & 50.03$\pm$0.01 & 42.38$\pm$0.33 & 47.88$\pm$0.02 & 53.01$\pm$0.08 & 42.75$\pm$0.11  \\
			\hline
			Video-ChatGPT \cite{maaz2023video} & S0 & 34.98$\pm$0.05 & 37.66$\pm$0.13 & 32.30$\pm$0.03 & 37.79$\pm$0.15 & 40.33$\pm$0.05 & 35.25$\pm$0.25  \\
			Video-ChatGPT \cite{maaz2023video} & S1 & 42.04$\pm$0.24 & 45.59$\pm$0.24 & 38.49$\pm$0.23 & 41.17$\pm$0.03 & 45.07$\pm$0.00 & 37.28$\pm$0.05  \\
			\rowcolor{lightgray}
			Video-ChatGPT \cite{maaz2023video} & S2 & 50.73$\pm$0.06 & 54.03$\pm$0.04 & 47.44$\pm$0.07 & 55.34$\pm$0.02 & 61.15$\pm$0.10 & 49.52$\pm$0.06  \\
			\hline
			Video-LLaMA \cite{zhang2023video} & S0 & 28.18$\pm$0.27 & 28.64$\pm$0.36 & 27.72$\pm$0.18 & 30.72$\pm$0.11 & 30.09$\pm$0.14 & 31.34$\pm$0.08  \\
			Video-LLaMA \cite{zhang2023video} & S1 & 34.48$\pm$0.16 & 35.82$\pm$0.20 & 33.15$\pm$0.11 & 34.05$\pm$0.24 & 35.16$\pm$0.22 & 32.94$\pm$0.26  \\
			\rowcolor{lightgray}
			Video-LLaMA \cite{zhang2023video} & S2 & 44.74$\pm$0.14 & 44.14$\pm$0.13 & 45.34$\pm$0.15 & 47.27$\pm$0.03 & 47.98$\pm$0.07 & 46.56$\pm$0.01  \\
			\hline
			VideoChat \cite{li2023videochat} & S0 & 31.95$\pm$0.01 & 31.73$\pm$0.13 & 32.17$\pm$0.10 & 34.56$\pm$0.02 & 33.53$\pm$0.01 & 35.60$\pm$0.05  \\
			VideoChat \cite{li2023videochat} & S1 & 45.13$\pm$0.07 & 46.24$\pm$0.05 & 44.01$\pm$0.10 & 44.25$\pm$0.09 & 44.76$\pm$0.02 & 43.75$\pm$0.16  \\
			\rowcolor{lightgray}
			VideoChat \cite{li2023videochat} & S2 & 45.70$\pm$0.09 & 42.90$\pm$0.27 & 48.49$\pm$0.10 & 45.63$\pm$0.04 & 47.20$\pm$0.12 & 44.05$\pm$0.05  \\
			\hline
			VideoChat2 \cite{li2024mvbench} & S0 & 36.78$\pm$0.04 & 43.08$\pm$0.00 & 30.47$\pm$0.09 & 36.01$\pm$0.01 & 41.16$\pm$0.00 & 30.86$\pm$0.01  \\
			VideoChat2 \cite{li2024mvbench} & S1 & 38.53$\pm$0.05 & 44.62$\pm$0.00 & 32.43$\pm$0.10 & 39.51$\pm$0.10 & 45.14$\pm$0.13 & 33.88$\pm$0.08  \\
			\rowcolor{lightgray}
			VideoChat2 \cite{li2024mvbench} & S2 & 49.60$\pm$0.28 & 54.72$\pm$0.41 & 44.47$\pm$0.15 & 49.90$\pm$0.06 & 57.12$\pm$0.08 & 42.68$\pm$0.04  \\
			\hline
			mPLUG-Owl \cite{ye2023mplug} & S0 & 39.25$\pm$0.14 & 40.56$\pm$0.15 & 37.94$\pm$0.12 & 40.53$\pm$0.33 & 40.44$\pm$0.24 & 40.62$\pm$0.43  \\
			mPLUG-Owl \cite{ye2023mplug} & S1 & 45.85$\pm$0.05 & 47.49$\pm$0.04 & 44.22$\pm$0.07 & 48.01$\pm$0.04 & 49.33$\pm$0.03 & 46.69$\pm$0.05  \\
			\rowcolor{lightgray}
			mPLUG-Owl \cite{ye2023mplug} & S2 & 52.79$\pm$0.13 & 54.54$\pm$0.13 & 51.04$\pm$0.13 & 51.43$\pm$0.03 & 56.40$\pm$0.11 & 46.47$\pm$0.18  \\
			\hline
			SALMONN \cite{tang2023salmonn} & S0 & 40.72$\pm$0.11 & 41.38$\pm$0.25 & 40.07$\pm$0.04 & 43.45$\pm$0.23 & 43.24$\pm$0.30 & 43.66$\pm$0.16  \\
			SALMONN \cite{tang2023salmonn} & S1 & 39.80$\pm$0.04 & 39.54$\pm$0.01 & 40.05$\pm$0.06 & 41.43$\pm$0.13 & 41.11$\pm$0.03 & 41.76$\pm$0.22  \\
			\rowcolor{lightgray}
			SALMONN \cite{tang2023salmonn} & S2 & 48.06$\pm$0.04 & 50.20$\pm$0.04 & 45.92$\pm$0.04 & 48.53$\pm$0.03 & 52.24$\pm$0.00 & 44.82$\pm$0.05  \\
			\hline
			Qwen-Audio \cite{chu2023qwen} & S0 & 33.03$\pm$0.04 & 41.92$\pm$0.00 & 24.14$\pm$0.08 & 32.59$\pm$0.08 & 40.84$\pm$0.13 & 24.33$\pm$0.03  \\
			Qwen-Audio \cite{chu2023qwen} & S1 & 37.49$\pm$0.11 & 46.69$\pm$0.15 & 28.29$\pm$0.08 & 46.81$\pm$0.00 & 58.08$\pm$0.00 & 35.53$\pm$0.00  \\
			\rowcolor{lightgray}
			Qwen-Audio \cite{chu2023qwen} & S2 & 40.23$\pm$0.09 & 49.42$\pm$0.18 & 31.04$\pm$0.00 & 43.53$\pm$0.04 & 53.71$\pm$0.00 & 33.34$\pm$0.09  \\
			\hline
			Video-LLaVA \cite{lin2023video} & S0 & 32.65$\pm$0.03 & 33.31$\pm$0.01 & 32.00$\pm$0.05 & 32.76$\pm$0.03 & 33.19$\pm$0.06 & 32.33$\pm$0.00  \\
			Video-LLaVA \cite{lin2023video} & S1 & 30.59$\pm$0.01 & 34.10$\pm$0.03 & 27.08$\pm$0.05 & 31.99$\pm$0.11 & 33.40$\pm$0.19 & 30.58$\pm$0.04  \\
			\rowcolor{lightgray}
			Video-LLaVA \cite{lin2023video} & S2 & 47.12$\pm$0.15 & 48.58$\pm$0.02 & 45.66$\pm$0.29 & 49.59$\pm$0.05 & 53.95$\pm$0.03 & 45.23$\pm$0.13  \\
			\hline
			LLaMA-VID \cite{li2023llama} & S0 & 35.20$\pm$0.14 & 36.71$\pm$0.15 & 33.69$\pm$0.14 & 33.30$\pm$0.04 & 33.12$\pm$0.06 & 33.48$\pm$0.03  \\
			LLaMA-VID \cite{li2023llama} & S1 & 42.43$\pm$0.03 & 43.97$\pm$0.04 & 40.89$\pm$0.03 & 42.57$\pm$0.08 & 43.28$\pm$0.11 & 41.86$\pm$0.04  \\
			\rowcolor{lightgray}
			LLaMA-VID \cite{li2023llama} & S2 & 51.29$\pm$0.09 & 52.71$\pm$0.18 & 49.87$\pm$0.00 & 52.45$\pm$0.02 & 57.30$\pm$0.00 & 47.61$\pm$0.03  \\
			\hline
			Chat-UniVi \cite{jin2023chat} & S0 & 40.02$\pm$0.18 & 42.32$\pm$0.21 & 37.72$\pm$0.15 & 36.85$\pm$0.30 & 37.74$\pm$0.27 & 35.96$\pm$0.33  \\
			Chat-UniVi \cite{jin2023chat} & S1 & 48.11$\pm$0.19 & 50.96$\pm$0.20 & 45.26$\pm$0.18 & 47.04$\pm$0.00 & 48.07$\pm$0.00 & 46.01$\pm$0.00  \\
			\rowcolor{lightgray}
			Chat-UniVi \cite{jin2023chat} & S2 & 53.09$\pm$0.01 & 53.68$\pm$0.00 & 52.50$\pm$0.02 & 54.20$\pm$0.02 & 58.54$\pm$0.01 & 49.86$\pm$0.03  \\
			\hline
		\end{tabular}
	}
\end{table}

\clearpage
\section{Metric Correlation Analysis}
\label{appendix_sec:metric}
\begin{figure*}[h]
	\centering
	\includegraphics[width=0.66\linewidth]{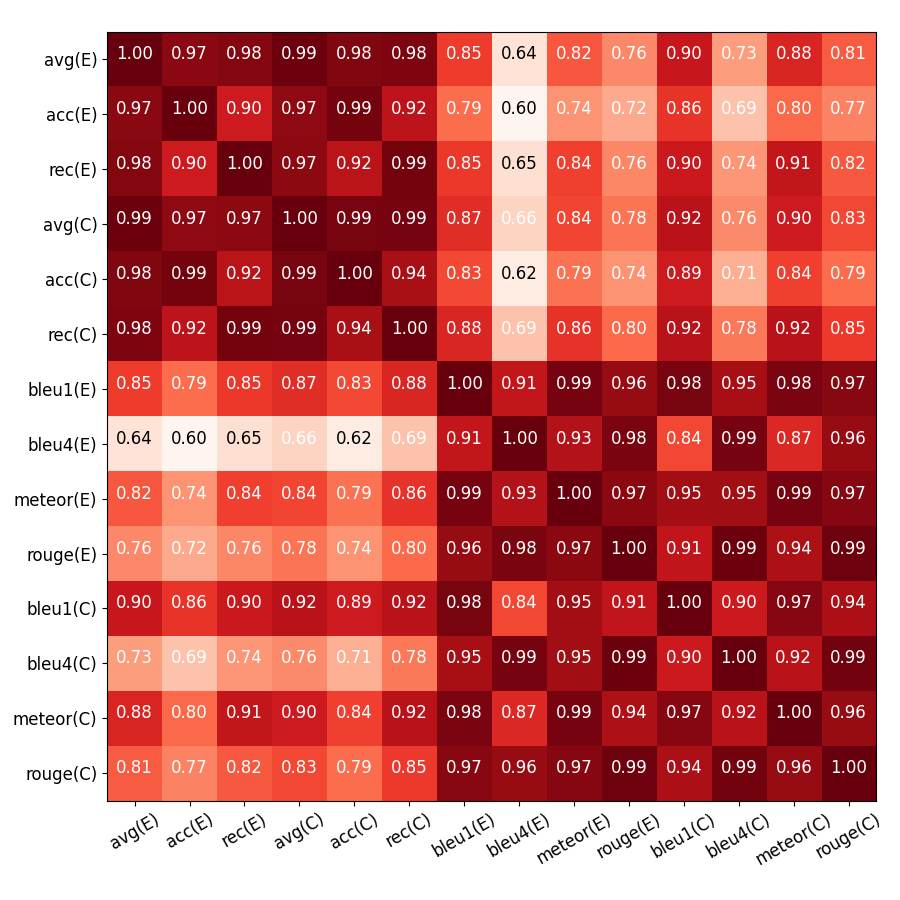}
	\caption{Visualization of metric correlations. In this figure, we consider both metric and language differences. Here, ``E'' and ``C'' represent English and Chinese, respectively.}
	\label{Figure8}
\end{figure*}

Table \ref{Table8} provides raw scores for matching-based metrics. In Figure \ref{Figure8}, we combine the results in Table \ref{Table3} and Table \ref{Table8} and calculate the PCC correlation scores between different metrics. In this figure, we consider emotion-based metrics  (i.e., Avg, $\mbox{Accuracy}_{\mbox{s}}$, $\mbox{Recall}_{\mbox{s}}$) and matching-based metrics (i.e., BLEU$_{1}$, BLEU$_{4}$, METEOR, ROUGE$_l$). Meanwhile, we consider the language influence.

\begin{table}[h]
	\centering
	\renewcommand\tabcolsep{3.8pt}
	\renewcommand\arraystretch{1.06}
	\caption{Performance of different models on matching-based metrics.}
	\label{Table8}
	\scalebox{0.8}{
		\begin{tabular}{lccc|cccc|cccc}
			\hline
			\multirow{2}{*}{Model} & \multirow{2}{*}{L} & \multirow{2}{*}{V} & \multirow{2}{*}{A} & \multicolumn{4}{c|}{English} & \multicolumn{4}{c}{Chinese} \\
			&&&&BLEU$_{1}$ &BLEU$_{4}$ &METEOR & ROUGE$_l$ &BLEU$_{1}$ &BLEU$_{4}$ &METEOR & ROUGE$_l$\\
			\hline
			Empty          & $\times$ & $\times$ & $\times$&0.00 & 0.00 & 0.67 & 1.75& 0.00 & 0.00 & 1.49 & 2.38\\
			Random         & $\times$ & $\times$ & $\times$&0.03 & 0.01 & 3.87 & 7.75& 0.01 & 0.00 & 3.18 & 5.98\\
			\hline
			Qwen-Audio \cite{chu2023qwen}   & $\surd$  & $\times$ & $\surd$&21.87 & 6.55 & 21.65 & 20.81& 27.64 & 12.07 & 26.09 & 25.24\\
			OneLLM \cite{han2023onellm}     & $\surd$  & $\times$ & $\surd$&33.81 & 8.54 & 28.00 & 22.46& 42.75 & 16.60 & 34.42 & 26.81\\
			Otter  \cite{li2023otter}   & $\surd$  & $\surd$ & $\times$&27.26 & 7.55 & 23.42 & 21.05& 35.35 & 14.41 & 29.34 & 25.91\\
			VideoChat  \cite{li2023videochat} & $\surd$  & $\surd$ & $\times$&26.44 & 5.41 & 30.58 & 19.11& 31.36 & 10.86 & 37.48 & 22.57\\
			Video-LLaMA \cite{zhang2023video}  & $\surd$ & $\surd$ & $\times$&28.76 & 6.41 & 31.22 & 20.41& 34.88 & 12.13 & 37.61 & 24.25\\
			PandaGPT \cite{su2023pandagpt}     & $\surd$  & $\surd$  & $\surd$&33.69 & 7.64 & 30.29 & 22.07& 43.02 & 15.83 & 37.94 & 26.87\\
			SALMONN \cite{tang2023salmonn}      & $\surd$  & $\times$ & $\surd$&31.89 & 7.19 & 28.42 & 20.99& 39.00 & 14.00 & 35.12 & 25.35\\
			Video-LLaVA \cite{lin2023video}   & $\surd$  & $\surd$ & $\times$&33.48 & 8.25 & 29.68 & 22.34& 42.72 & 15.97 & 36.87 & 26.90\\
			VideoChat2 \cite{li2024mvbench}   & $\surd$  & $\surd$ & $\times$&31.60 & 8.10 & 26.61 & 21.65& 41.18 & 16.15 & 33.54 & 26.80\\
			OneLLM \cite{han2023onellm}     & $\surd$  & $\surd$  & $\times$&32.19 & 8.10 & 28.44 & 22.25& 41.31 & 15.15 & 35.15 & 25.98\\
			LLaMA-VID \cite{li2023llama}    & $\surd$  & $\surd$  & $\times$&33.81 & 8.26 & 30.31 & 22.36& 43.01 & 16.23 & 37.92 & 27.20\\
			mPLUG-Owl \cite{ye2023mplug}   & $\surd$  & $\surd$ & $\times$&33.04 & 7.75 & 30.24 & 21.75& 41.69 & 15.16 & 37.81 & 26.39\\
			Video-ChatGPT \cite{maaz2023video} & $\surd$ & $\surd$ & $\times$&32.64 & 7.65 & 30.25 & 22.01& 41.96 & 15.50 & 38.18 & 26.35\\
			Chat-UniVi \cite{jin2023chat}   & $\surd$  & $\surd$  & $\times$&32.80 & 7.83 & 31.12 & 22.15& 40.76 & 15.05 & 38.75 & 26.43\\
			GPT-4V \cite{openai2023gpt4v}     & $\surd$   & $\surd$ & $\times$&39.40 & 18.41 & 43.67 & 32.60& 45.45 & 29.08 & 53.76 & 40.37\\
			\hline
			EMER(Text)  & $\surd$  & $\times$ & $\times$&18.97 & 5.32 & 18.55 & 16.93& 25.24 & 10.30 & 23.01 & 20.15\\
			EMER(Video) & $\times$ & $\surd$  & $\times$&48.31 & 30.35 & 41.93 & 42.69& 58.19 & 42.65 & 51.86 & 49.36\\
			EMER(Audio) & $\surd$  & $\times$ & $\surd$&34.19 & 17.54 & 30.86 & 32.87& 46.74 & 30.80 & 42.19 & 40.24\\
			EMER(Multi) & $\surd$  & $\surd$  & $\surd$&100.0 & 100.0 & 100.0 & 100.0& 100.0 & 100.0 & 100.0 & 100.0\\
			\hline
		\end{tabular}
	}
\end{table}

\end{document}